\documentclass[reprint]{revtex4-1}
\usepackage{graphicx}
\usepackage{amssymb}
\usepackage{amsmath,amsfonts,amsthm} 
\usepackage{mhchem}
\usepackage{dsfont}

\begin{document}

\title{Energy and entropy on irreversible chemical reaction-diffusion systems with asymptotic stability}

\author{Aldo Ledesma-Dur\'an}
\email[]{aldo\_ledesma@xanum.uam.mx}
\address{Departamento de Matem\'aticas,
Universidad Aut\'onoma Metropolitana
Iztapalapa, Ciudad de M\'exico, Mexico}

\author{Iv\'an Santamar\'ia Holek }
\address{Unidad Multidisciplinaria de Docencia e Investigaci\'on, Universidad   Nacional Aut\'{o}noma de M\'{e}xico. Boulevard Juriquilla 3001,   Juriquilla, 76230 Quer\'etaro, Mexico.}

\vspace{10pt}


\begin{abstract}
In the framework of irreversible thermodynamics, we study autonomous systems of reaction-diffusion equations to show how the entropy and free energy of an open and irreversible reactor depend on concentrations. To do this, we find a Lyapunov function that depends directly on the eigenvalues and eigenvectors of the linearized problem  valid for linear and asymptotically stable systems. By suggesting the role of this Lyapunov function as internal free energy, we were able to reflect the properties of the dynamical system directly into the second law of thermodynamics making a redefinition of the chemical potentials. We demonstrate the consistency of our hypotheses with basic thermodynamic principles such as the proportionality between flows and forces proposed by Onsager, the spectral decomposition of entropy production and the Glansdorff-Prigogine evolution criterion of a thermodynamic system.
\end{abstract}

\maketitle
%
\vspace{2pc}
%
%
%
%
\section{Introduction}
 \label{sec:introduction}
 
Reaction-Diffusion (RD) equations constitute a standard model for spatial and temporal organization \cite{murray2007mathematical,murray2001mathematical, keener1998mathematical, cross2009pattern}. In a closed system, the solution is expected to maximize entropy or minimize free energy \cite{sandler2017chemical, eu2018chemical}. However, for many cases of practical importance, the reactor is an open system where the chemostat species are continuously pumped into or out of the reactor \cite{qian2006open,sethna2006statistical}.  In addition to these opening conditions, the irreversibility of many processes is generally coupled with non-equilibrium conditions where external flows can give rise to spatial inhomogeneities and dissipative structures such as Turing patterns or traveling waves \cite{prigogine1967symmetry,mori2013dissipative,avanzini2019thermodynamics}. The existence of these dissipative structures is based on the incessant entry of matter or energy into the system allowing the dissipative process, instead of bringing the components to the homogeneous spatial state, to lead to the organization of the agents \cite{kondepudi2014modern,feinberg1974dynamics}. In these circumstances, entropy production does not necessarily tend to zero as in an isolated system and its evolution over time must be linked to the evolution of the trajectories in the phase space of the solutions \cite{rao2016nonequilibrium, desvillettes2017trend}.

The irreversible thermodynamic formalism allows us to find entropy production when chemical exchange occurs primarily through a series of reversible reactions. However, this is not a typical situation, and there are only a few case studies in which the irreversibility of reactions has been considered \cite {desvillettes2007entropy, esposito2020open,fischer2015global,gentil2009asymptotic}.  For example, under the assumption of reversibility in chemical reactions, the detailed balance condition allows us to find an equilibrium of fluxes by canceling the affinities. However, in the case of irreversible reactions, the evolution towards the stationary solution (the one that no longer changes with time) follows a dynamic given by the stability of the fixed points of the dynamic system. In this sense, the conditions for the equilibrium of a reversible thermodynamic system are more general than the dynamic stability of a fixed point solution for the reversible case \cite{fischer2017weak,daus2019entropic,lepoutre2017entropic}. 
This distinction is fundamental because if the flows and forces defined for reversible reactions are used in irreversible reaction-diffusion problems, then the entropy production is not necessarily positive or well defined.  It turns out that the solutions that are stable in the context of dynamic systems (and that are perfectly observable in the modeling of actual processes) are not valid in the thermodynamical frame. Therefore, the dynamic concept of stability of solutions (negative real part of all eigenvalues) is not necessarily the same as the positivity of the proportionality matrix between flows and forces when the classical definitions of chemical potential are used in the known scheme of irreversible thermodynamics \cite{fellner2017entropy,plastino2000nonextensive}. 
Therefore, classical schemes for quantifying Gibbs energy, entropy production rate, and chemical potentials are restricted to reversible systems. And although it seems true that all systems are, to some extent, reversible \cite{gorban2011extended}, most of the dynamic systems frequently used in the modeling of biological, social and physical systems are not, making the condition of detailed balance too restrictive
\cite{gnesotto2018broken,ge2012landscapes,qian2013decomposition}.

In this work, we use the theory of irreversible thermodynamics to study the entropy and free energy of an open and irreversible reaction-diffusion system \cite{kondepudi2014modern, nicolis1979irreversible}. The hypothesis is that the principles of maximization of entropy and minimization of energy reflect the asymptotic stability of the dynamical system. To make this connection, we will show that for irreversible process, a reinterpretation of chemical potentials is necessary, as well as the coupling between chemical species and diffusive fluxes \cite{heimburg2017linear, wang2019complex}. The bridge between thermodynamic principles and the concepts used in the study of dynamical systems will be a Lyapunov function that is a measure of the Gibbs free energy due to irreversible processes within the reactor \cite{nicolis1996nonequilibrium, di2008entropy, lepri1997entropy, desvillettes2008entropy}.

In this work, we only consider systems that are asymptotically stable under the linearization \cite{michaelian2002non,fellner2015entropy}. For this class of systems, we find a simple form of the Lyapunov potential by using the proportionality between forces and flows. Since we are considering linear systems, we can also focus on a single fixed point. For nonlinear conditions, if a system has more than one fixed point, our definitions of entropy and energy will be valid only locally in the neighborhood of each point. For systems whose fixed point is linearly unstable but which result in auto-organizing solutions such as Turing patterns, waves, and limit cycles, we will detail in a future work how to find the energy and entropy functions in terms of the approximate solutions of the Ginzburg-Landau approach \cite{ledesmafuturo}.

\section{Linearly stable reactions}

Let us start, as  introduction to the problem, by considering a general reaction-diffusion system, at constant pressure and temperature, that can be represented by the following set of equations
\begin{equation}\label{eq:rd}
\frac{d \mathbf{n}}{ dt }=\mathds{D} \nabla^2 \mathbf{n} +\mathbf{F(\mathbf{n})}.
\end{equation}

\noindent where $\mathbf{n}$ is a column vector with the concentration of the chemical components and  $\mathbf{F}$ are the reactions  that, for most of the processes,  are usually non-linear functions. Here, $\mathds{D}$ is the matrix of diffusion coefficients and $\nabla^2$ is the Laplacian operator. In this equation, concentrations and parameters are real-valued quantities.

In the proximity of the fixed point $\mathbf{n}_0$ (\emph{i.e.} for initial conditions near the stationary concentrations), where $\mathbf{F}(\mathbf{n}_{0})=\mathbf{0}$, each function $\mathbf{F}_k$ 
can be expanded in terms of its Taylor series. Introducing the notation 
$\mathbf{u}=\mathbf{n}-\mathbf{n}_0$ for the deviation with respect to the stationary concentration, we can write 
\begin{equation}\label{eq:expanion}
\frac{d\mathbf{u}}{dt}=\mathds{D} \nabla^2 \mathbf{u}+\mathds{J} \mathbf{u}+\mathds{M} \mathbf{u}\mathbf{u}+\mathds{N} \mathbf{u}\mathbf{u}\mathbf{u}+\cdots,
\end{equation}

\noindent where $\mathds{J}$ is the Jacobian at $\mathbf{n}_0$ and the higher order terms are 

\begin{equation}
(\mathds{M}\mathbf{u}\mathbf{u})_i=\frac{1}{2!}\sum\limits_{j,k} \frac{\partial^2 F_i(\mathbf{n}_0)}{ \partial n_j \partial n_k} \mathbf{u}_j \mathbf{u}_k,
\end{equation}

\begin{equation}
(\mathds{N}\mathbf{u}\mathbf{u}\mathbf{u})_i=\frac{1}{3!}\sum\limits_{j,k,l} \frac{\partial^3 F_i(\mathbf{n}_0)}{ \partial n_j \partial n_k\partial n_l} \mathbf{u}_j \mathbf{u}_k \mathbf{u}_l.
\end{equation}
For the sake of clarity in the exposition of the physical formalism, we will focus on the linear approximation, leaving for future work the consideration of weakly nonlinearities. Therefore in this work we consider those linear systems and non-linear systems that are asymptotically stables; in the first case with exact global Lyapunov functions and, in the second case, only as a local approximation valid near a specific fixed point. Non-linear systems where the fixed point is unstable and resulting in dissipative structures are not included here \cite{ledesmafuturo}.

\subsection{Linear chemical reactions}
Let us start by considering reactive systems that are linearly stable. This means that its dynamics is determined mainly by the linear terms of the reaction in the expansion (\ref{eq:expanion}), therefore

\begin{equation}\label{eq:linear}
\frac{d \mathbf{u}}{ dt } \cong \mathds{J} \mathbf{u},
\end{equation}

\noindent for the $N$ chemical components. We define a new set of coordinates $\mathbf{y}$ through the transformation

\begin{equation}\label{eq:change}
\mathbf{u}=\mathds{P} \mathbf{y},
\end{equation}
\noindent where $\mathds{P}=[ \mathbf{v}_1 \, \mathbf{v}_1 \cdots  \mathbf{v}_N ]$ is the square matrix where each column correspond to an eigenvector of $\mathds{J}$. From this,  $\mathds{J}  \mathbf{v}_k=\lambda_k  \mathbf{v}_k$, where $\lambda_k$ are the eigenvalues.  We will assume  that all eigenvalues have non-positive real part in order to consider asymptotic stability. Besides, for mathematical simplicity, the eigenvectors are assumed linearly independent guaranteeing the existence of $\mathds{P}^{-1}$. The change of variables allows us to deduce from  (\ref{eq:linear}) that:

\begin{equation}\label{eq:diagonal}
\frac{d \mathbf{y}}{ dt } = \mathds{P}^{-1}  \mathds{J}  \mathds{P}  \mathbf{y} =\Lambda^{(J)} \mathbf{y},
\end{equation}
\noindent where  $\{\Lambda^{J}\}_{j\,k}= \lambda_j \delta_{jk}$ is a diagonal matrix whose entries correspond to the eigenvalues of $\mathds{J}$. The superscript $(J)$ emphasizes that this matrix considers only the linear part of the chemical reaction. Here $\delta_{jk}$ is the Kronecker delta. The change of variables allows us to uncouple each component coordinate and have

\begin{equation}\label{eq:particulares}
\frac{d y_j}{dt}=\lambda_j y_j  \Rightarrow y_k(t)=c_j e^{\lambda_j t},
\end{equation}
\noindent where $c_k$ are complex constants that depend on the initial conditions.

\subsection{Lyapunov function}

Now, let us define the following  candidate to Lyapunov function $\mathcal{V}(t)$  through its spatial derivative as

\begin{equation}\label{eq:lyapunov}
\frac{1}{V}\frac{d \mathcal{V}}{d t} (\mathbf{u}) = \hat{ \boldsymbol{\mu}}^T \frac{d \mathbf{u}}{ dt },
\end{equation}
\noindent where  $\hat{\boldsymbol{\mu}}$ is an dimensionless chemical potential that we find later and the superscript $T$ is the transpose symbol. Here $V$ is the volume of the domain that is assumed constant. Substituting (\ref{eq:linear})  and (\ref{eq:change}) in (\ref{eq:lyapunov}), we have  $V^{-1}d\mathcal{V}/dt= \hat{\boldsymbol{\mu}}^T \mathds{J} \mathds{P} \mathbf{y}$. To prove that $\mathcal{V}$ is a Lyapunov function, given the uncoupling in  (\ref{eq:diagonal}), we propose that the dimensionless chemical potential at first order has the form
 
\begin{equation}\label{eq:fluxesforces}
\hat{\boldsymbol{\mu}}= \mathds{P}^{-T} \overline{\mathbf{y}}.
\end{equation} 
\noindent The overline represents complex conjugate. The reason for this hypothesis will be explained below. Using this equation and the diagonalization in (\ref{eq:diagonal}) we have:

\begin{equation}\label{eq:sumadvt}
\frac{1}{V} \frac{d \mathcal{V}}{d t}  = \overline{\mathbf{y}}^T \Lambda^{J} \mathbf{y} =\sum \limits_{k=1}^N   \lambda_k |y_k|^2.
\end{equation}
\noindent  The $|\cdot|$ symbol refers to the complex norm and $y_k$ is the $k$ component of the vector $\mathbf{y}$. For clarity, we can separate this sum in three terms which correspond to real solutions with real eigenvalues $\lambda_r=r$, spiral solutions with $\lambda_s=p+i q$ with $p\neq 0$ and $i=\sqrt{-1}$, and those with eigenvalues with null real part, $Re\{\lambda_b\}=0$. Under this separation, the sum in (\ref{eq:sumadvt}) can be split as 

\begin{equation}\label{eq:Vtimederivative}
\frac{1}{V} \frac{d \mathcal{V}}{d t}  =\sum \limits_{r=1}^{R}  \lambda_r y_r^2 +  \sum \limits_{s=R+1}^{S+R}  Re \{\lambda_s\} |y_s|^2 
\end{equation}
where $R,S$ are the number of real and complex eigenvalues, respectively and $B=N-R-S$ are those with null real part and  that do no contribute to the time change of $\mathcal{V}$. In the last equation, as well as in 
(\ref{eq:fluxesforces}), it is important to remember that complex solutions  $y_k$ of  systems of equations of real coefficients  always come in pairs of complex conjugates.

Equation (\ref{eq:Vtimederivative}) shows that the time derivative of $\mathcal{V}$ is non-positive as long as the real part of all eigenvalues is non-positive, and attaches  its minimum value at 0 when all displacements $|y_k|$ are null. Therefore, we can conclude that for linear stable systems, 

\begin{equation}
\frac{d \mathcal{V}}{dt}\leq 0.
\end{equation}  

In order to find the value of  $\mathcal{V}(t)$, we integrate Eq. (\ref{eq:Vtimederivative})  from time $t$ to $\infty$. Using Eq. (\ref{eq:particulares}) we obtain

\begin{equation}\label{eq:Vtimederivative14}
\mathcal{V} (t)-\mathcal{V}_{\infty} =\frac{V}{2} \left[  \sum \limits_{r=1}^{R}   y_r^2 +  \sum \limits_{s=R+1}^{S+R} |y_s|^2  \right],
\end{equation}
\noindent where $\mathcal{V}_{\infty}$ is the reference value of $\mathcal{V}$ in the stationary case when $t \to \infty$.  Since the eigenvalues with null real part do no contribute  to the time derivative of $\mathcal{V}$, we can add these terms to the solution of $\mathcal{V}$ as the additive constant $\mathcal{V}_{\infty} =(V/2)\sum_b |y_b|^2$ and obtain from (\ref{eq:change}) that

\begin{equation}\label{eq:lyapireacti}
\mathcal{V}(t)=\frac{V}{2}|\mathbf{y}|^2=\frac{1}{2}\int_V   |\mathds{P}^{-1} (\mathbf{n}-\mathbf{n}_0)|^2 \, dv.
\end{equation}
From this equation it is clear that $\mathcal{V}\geq 0$ and is only 0 when $\mathbf{n}=\mathbf{n}_0$ since the nullspace of an invertible matrix is only the origin. From here, it is deduced that $\mathcal{V}$ is a Lyapunov function for the linearized system in (\ref{eq:linear}) for the stable point $\mathbf{n}_0$.

\subsection{Entropy production and Lyapunov function}

Based on the above results, it is plausible to postulate the following equality between the entropy production $\sigma $ and the Lyapunov function (\ref{eq:lyapunov}):
\begin{equation}\label{eq:entropy}
\int _V\sigma d\,v = -\int _V \frac{1}{T} \Delta \boldsymbol{\mu}^T \frac{d \mathbf{n}}{ dt } d\,v   \equiv -\mathcal{R} \frac{d \mathcal{V}}{d t},
\end{equation}
where $\mathcal{R}$ is the gas constant and $T$ the temperature. 
Substitution of the time derivative of Eq. (\ref{eq:lyapireacti}) into (\ref{eq:entropy}) yields 
\begin{equation}\label{eq:entropy-explicit2}
\sigma  = -\mathcal{R} (\mathbf{n}-\mathbf{n}_0)^T  (\mathds{L}^{J})^T \frac{d\mathbf{n}}{ dt } ,
\end{equation}
where the matrix
\begin{equation}\label{eq:lj}
\mathds{L}^{J}=\mathds{P}^{-T} \overline{\mathds{P}^{-1}},
\end{equation}
can be identified with the matrix of phenomenological Onsager coefficients. Eq. (\ref{eq:lj}) allows us to identify the chemical potential as 
\begin{equation}\label{eq:entropy-explicit2}
\Delta \boldsymbol{\mu} = \mathcal{R}T \mathds{L}^{J} (\mathbf{n}-\mathbf{n}_0), 
\end{equation}
\noindent relation which expresses the proportionality between  fluxes and forces  of the linear theory \cite{onsager1953fluctuations,machlup1953fluctuations,kondepudi2014modern}, valid near the fixed point concentrations, $\mathbf{n}_0$.

Using the thermodynamic relationship between the free Gibbs energy $ G $ and the entropy $ S $ for a system with constant pressure, volume and temperature \cite{kondepudi2014modern}, $ \Delta G = -T \Delta S $, we find that close from the fixed point, the Gibbs energy differential due to \emph{internal} changes in the system (emphasized by the subscript $ i $) \cite{nieto2003entropy} is

\begin{equation} \label{eq:gibbs}
\Delta_i G=\mathcal{R} T \mathcal{V}=\frac{\mathcal{R} T}{2}|\mathds{P}^{-1} (\mathbf{n}-\mathbf{n}_0)|^2,
\end{equation}
\noindent where we have used $ \sigma = d_is / dt $. Therefore, near a fixed point concentration, the internal free energy has an upward concave parabolic profile as a minimum that reflects the steady-state and its stability properties. This means that if fluctuations disturb the system near that state, the system spontaneously returns to it and, therefore, the concept of stability in the thermodynamic sense is equivalent its stability as a dynamical system.

 It is not difficult to show that  $\mathds{L}^{J}$ in \eqref{eq:lj} is an Hermitian matrix since can be written as 
$\mathds{L}^{J} =\mathds{B} \overline{ \mathds{B}}^{T}$ with $\mathds{B}=\mathds{P}^{-T} $ and, since this square matrix has full rank, then, $\mathds{L}^{J}$ is a positive definite matrix. The Hermitian nature of $\mathds{L}^{J}$ implies that it is possible to find a diagonalization of the matrix with only real positive entries $\ell_k$. This fact allows one to write the chemical potential of the k-esim species in its usual non-coupled form
\begin{equation}\label{mu-uncoupled}
 \Delta \tilde{\mu}_k \approx \mathcal{R} T \ell_k (\tilde{n}_k -\tilde{n}_{k0}). 
\end{equation}
The analysis of these formal aspects and other details concerning the mathematical structure of the space of solutions when   eigenvalues or eigenvectors are repeated, are beyond the scope of this work and will be studied in future works.

\begin{figure*}[btp]
\centering
\includegraphics[scale=0.15]{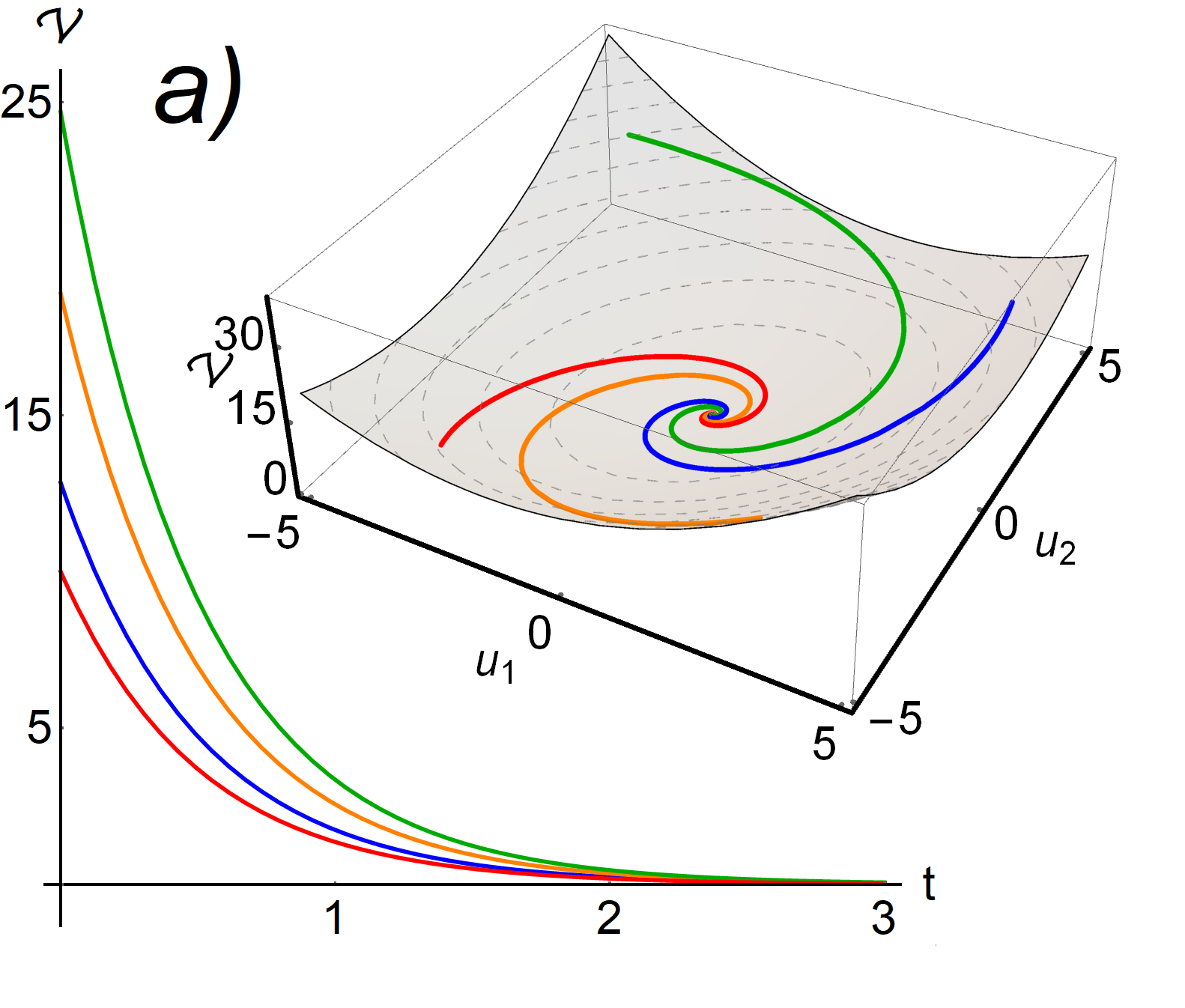}
\includegraphics[scale=0.12]{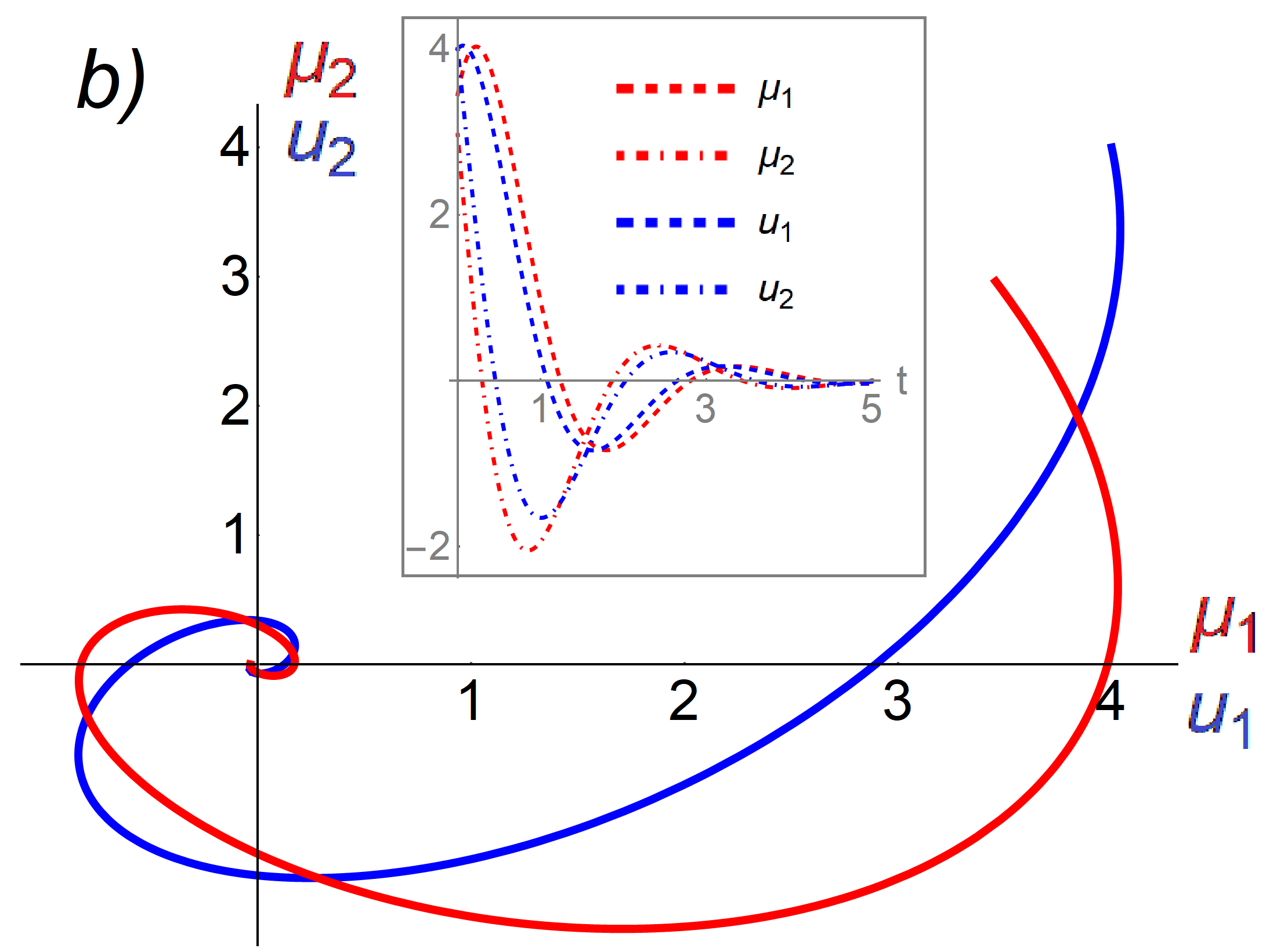}
\includegraphics[scale=0.22]{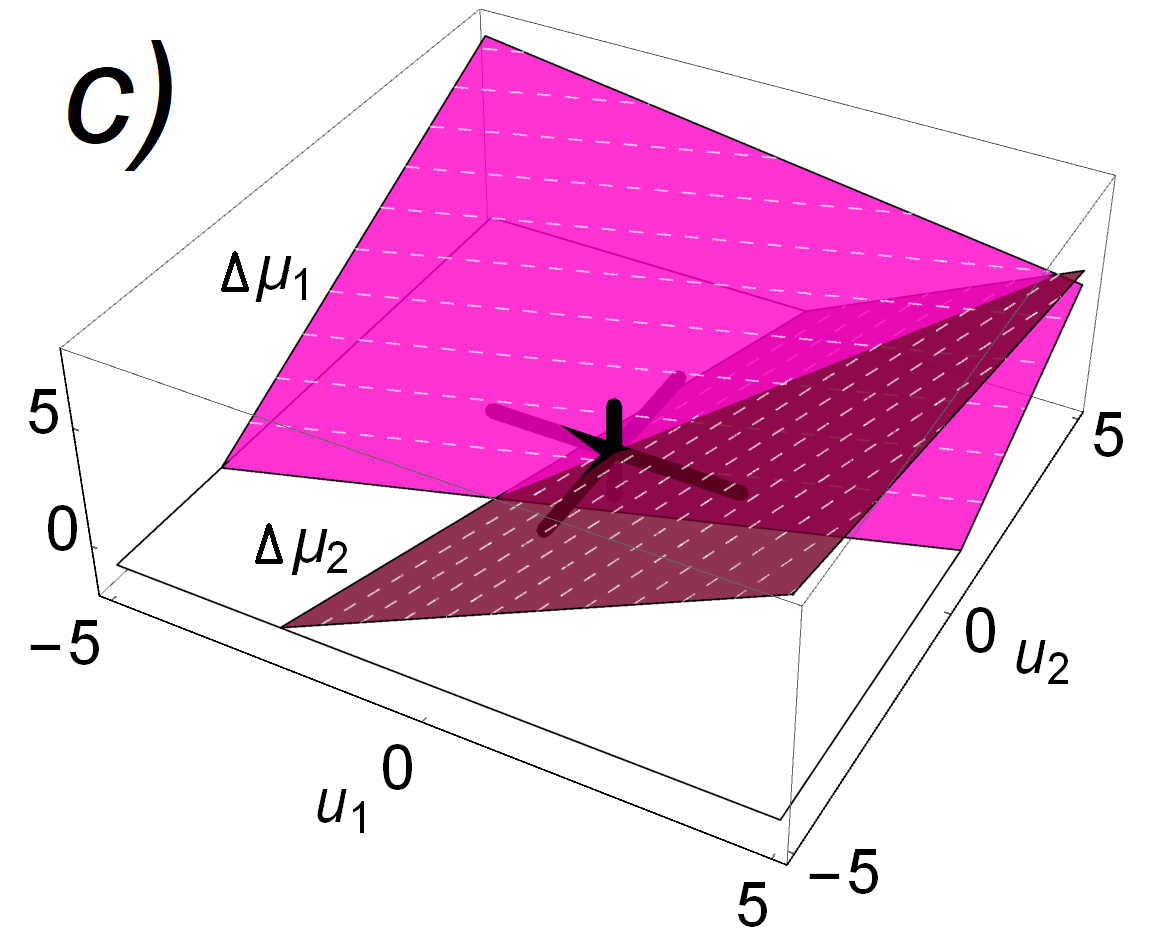}
\caption{Lyapunov function for a reactive system. a) $ \mathcal {V} $ as a function of time and the coordinates for some trajectories. b) Phase plane of concentrations and chemical potentials near the fixed point and their temporal behavior. c) Coupling between the chemical potentials in terms of differences of concentrations with respect to the fixed point.
}\label{fig:reaction}
\end{figure*}

Fig. \ref {fig:reaction} illustrates the above results. The system \eqref{eq:linear}, with the Jacobian matrix \eqref{eq:jota} exhibits oscillating solutions. The decay in time of the Lyapunov function $ \mathcal {V} $ is plotted in Fig. \ref{fig:reaction}a for four different initial conditions. The inset shows how the trajectories in the phase plane converge to the fixed point, where the concentrations of all components reach the constant concentration. The phase space of the concentrations and the chemical potentials in blue and red colors, respectively, are represented in Fig. \ref{fig:reaction}b. The inset shows the coordination between the concentrations and the oscillations of the chemical potential. They evolve so that the entropy production remains positive and reaches zero at the fixed point. Finally, Fig. \ref{fig:reaction}c shows that, near the fixed point, the chemical potentials are coupled and, therefore, their planes do not intersect on the axes $ u_1 $ or $ u_2 $ but on the principal directions of $ \mathds{L}^ {J}$.

\section{Purely diffusive systems}
Let us consider now a system of diffusive  non-reactive components of the form
\begin{equation}\label{eq:difusa}
\frac{d \mathbf{u}}{ dt }= \mathds{D}\nabla^2 \mathbf{u},
\end{equation}
where, for the sake of simplicity, the matrix of diffusion coefficients, $\mathds{D}$, is assumed of full rank. The initial condition is $\mathbf{n}(x,t=0)=\mathbf{n}^{ini}(x)$ and, just for illustrative purposes, we consider  a one-dimensional domain of length $V$ with homogeneous Dirichlet boundary conditions, $\mathbf{u}(x=0,t)=\mathbf{u}(x=V,t)=\mathbf{0}$.

Let us consider one solution of  equation \eqref{eq:diagonal}, \emph{i.e.} one eigenfunction with  wavenumber $\kappa_m$ of the diffusion operator:

\begin{equation}\label{eq:change2}
\mathbf{u}^{(m)}=\mathds{Q} \, \mathbf{y}^{(m)}(t) \, \sin ( \kappa_m  x),
\end{equation}
\noindent where the superscript $(m)$ numbers the eigenfunctions and  $\mathds{Q}$ is the matrix whose columns correspond to the eigenvectors of $\mathds{D}$. It is deduced that \eqref{eq:change2} is solution of \eqref{eq:difusa} if

\begin{equation}
\frac{d\mathbf{y}^{(m)}}{dt}=-\kappa^2_m \mathds{Q}^{-1} \mathds{D}  \mathds{Q} \mathbf{y}^{(m)}(t) = -\kappa_m^2 \Lambda^{D} \mathbf{y}^{(m)},
\end{equation}
where  $\{\Lambda^{D}\}_{j\,k}=d_j \delta_{jk}$ is a diagonal matrix with entries $d_j\geq 0$ that correspond to the non negative real eigenvalues of $\mathds{D}$. These equations have uncoupled solutions for each component as:
\begin{equation}\label{eq:ydifusa}
y^{m}_j(t) =c_j^m e^{-\kappa^2_m d_j t },
\end{equation}
\noindent where $c_j^m$ are integration constants whose meaning we discuss below.

Since the solution $\mathbf{u}(x,t)$ must fulfill boundary conditions,  the wavenumber has discrete values $\kappa_m= 2 m \pi /V$  with $m=1,2,3, \ldots$. Furthermore, the fulfillment of the initial conditions requires that the solution be a combination of the complete set of eigenfunctions numbered by $ m $ in the form:
\begin{equation}\label{eq:udifusa}
\mathbf{u}=\sum\limits_{m=1}^{\infty} \mathbf{u}^{(m)}(x,t).
\end{equation}
Using \eqref{eq:change2} and \eqref{eq:ydifusa} in this equation, it is clear that 
\begin{equation}
\left[\mathds{Q}^{-1}\mathbf{u}^{ini}(x) \right]_j=\sum\limits_{m=1}^{\infty} c_j^m \sin ( \kappa_m  x),
\end{equation}
and therefore, $c_j^m$ is the $m$-esim Fourier coefficient of the of the $j$-esim entry of  
$\mathds{Q}^{-1}\mathbf{u}^{ini}(x)$.

Now, we can define a dimensionless chemical potential for each wavenumber $\kappa_m$ of the form $\hat{\boldsymbol{\mu}}^{(m)}= \mathds{Q}^{-T} \bar{\mathbf{y}}^{(m)} \sin (\kappa_m x)$ and, the total   chemical potential as the sum of the chemical potentials for each mode:

\begin{equation}\label{eq:mudifusa}
\hat{\boldsymbol{\mu}}=\sum\limits_{m=1}^{\infty} \hat{\boldsymbol{\mu}}^{(m)}.
\end{equation}
\noindent In this case we can generalize the definition of the function $\mathcal{V}$ in \eqref{eq:lyapunov} for include spatially inhomogeneous systems as

\begin{equation}\label{eq:lyapunov2}
\frac{d \mathcal{V}}{d t} (t) =\int_V \hat{ \boldsymbol{\mu}}^T \frac{d \mathbf{u}}{ dt } \,dv,
\end{equation}
\noindent where $dv$ is the volume differential. Substituting \eqref{eq:fluxesforces}, \eqref{eq:udifusa}
and \eqref{eq:mudifusa} in this equation and using the orthogonality of the eigenfunctions

\begin{equation}\label{eq:ortogonal}
\int_0^V \sin(\kappa_j x)  \sin(\kappa_k x)   dv=(V/2) \delta_{jk},
\end{equation}
for $j,k=1,2,\ldots$ we have 

\begin{equation}
\frac{1}{V}\frac{d \mathcal{V}}{dt}=-\sum \limits_{m=1}^{\infty}   \kappa_m^2 \left( \bar{\mathbf{y}}^{(m)T} \Lambda ^{D}  \mathbf{y}^{(m)} \right) .
\end{equation}
\noindent This shows that $d\mathcal{V}/dt \leq 0$ given that the eigenvalues of $\mathds{D}$ are assumed real and positive. To see this, we write this equation in component form as:

\begin{equation}\label{eq:vparadifuso}
\frac{1}{V}\frac{d \mathcal{V}}{dt}=-\sum \limits_{m=1}^{\infty} \kappa_m^2  \sum \limits_{j=1}^{N} d_j \left|y_j^{(m)}\right|^2 .
\end{equation}
\noindent This equation is readily integrable from $t$ to $\infty$ using \eqref{eq:ydifusa} for obtaining 

\begin{equation}\label{eq:vdifusa1}
\mathcal{V}=\frac{V}{4} \sum\limits_{m=1}^{\infty} |\mathbf{y}^{(m)}|^2.
\end{equation}
Here, we have used the reference value $\mathcal{V}(\infty)$ for including possible null eigenvalues, where $d_j=0$ or spatially homogeneous modes $k_m$ which, as it occurs in \eqref{eq:vparadifuso}, represent only additive constants  that do not modify the value of the derivative of $\mathcal{V}(t)$. 

In terms of the original variables, using \eqref{eq:change2}, \eqref{eq:udifusa} and  \eqref{eq:ortogonal},  we have 

\begin{equation}\label{eq:lyapidif}
\mathcal{V}(t) =\frac{1}{2}   \int_V |  \mathds{Q}^{-1}  \mathbf{u}   |^2 dv.
\end{equation}
\noindent From this equation it is deduced that $V(t)\geq 0$, that $V(t)$ is only null when $\mathbf{u}=0$ and $d\mathcal{V}/dt \leq 0$. Therefore $\mathcal{V}$ in \eqref{eq:lyapidif} is a Lyapunov function for the solutions of the linear diffusion equation in \eqref{eq:difusa}.

Let us remark that Eq. \eqref{eq:lyapidif} shows the stability of the origin $\mathbf{u}=\mathbf{0}$ since, for the used boundary conditions (homogeneous Dirichlet), it corresponds to the stationary solution $\mathbf{u}^{ss}=\mathbf{0}$. However, the dependence of the Lyapunov function in \eqref{eq:lyapidif} with the stationary solution is more general and can also include non-equilibrium situations, as we will show at the end of this section.

Eq. \eqref{eq:entropy} establishes the correspondence between the entropy production per unit time and the Lyapunov function. According to  Eq. \eqref{eq:gibbs}, the chemical potential for the diffusive process is given now by 
\begin{equation}\label{eq:potentialdifuso}
\Delta\boldsymbol{\mu} \cong \mathcal{R} T \hat{\boldsymbol{\mu}}= \mathcal{R} T \mathds{L}^{D} (\mathbf{n}-\mathbf{n}^{ss}),
\end{equation}
\noindent where 
\begin{equation}\label{eq:LD}
\mathds{L}^{D}=\mathds{Q}^{-T} \overline{\mathds{Q}^{-1}},
\end{equation}
\noindent  is the positive definite matrix of proportionality between forces and fluxes near the stationary state for a purely diffusive process. Finally, we must point out that our formalism would apply in the same way if the eigenvalues of the diffusion matrix could be complex numbers as long as their real part is positive. However, this case seems physically unreal, as discussed in Ref. \cite{vanag2009cross}.

\begin{figure*}[btp]
\centering
\includegraphics[scale=0.12]{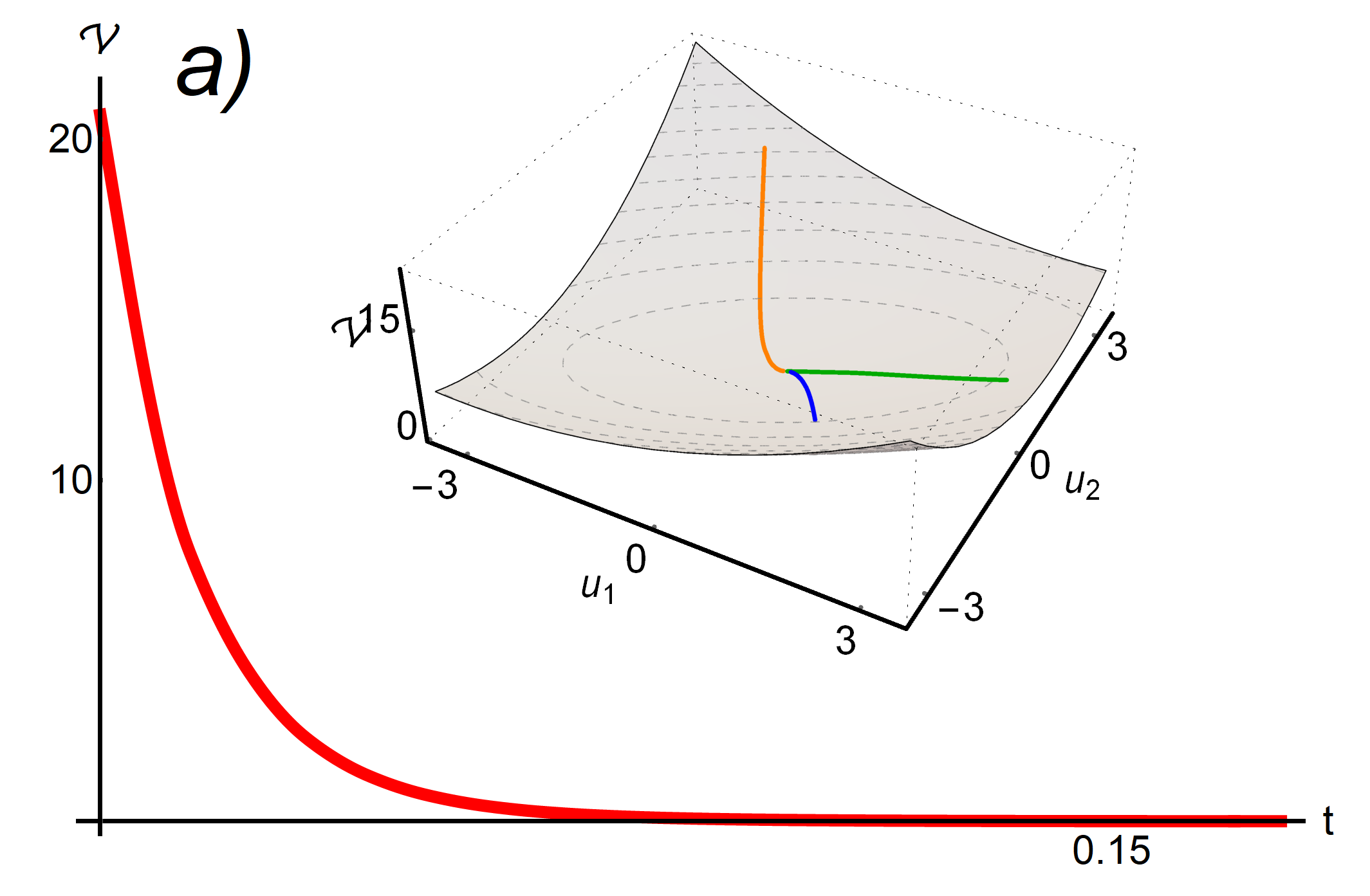}
\includegraphics[scale=0.15]{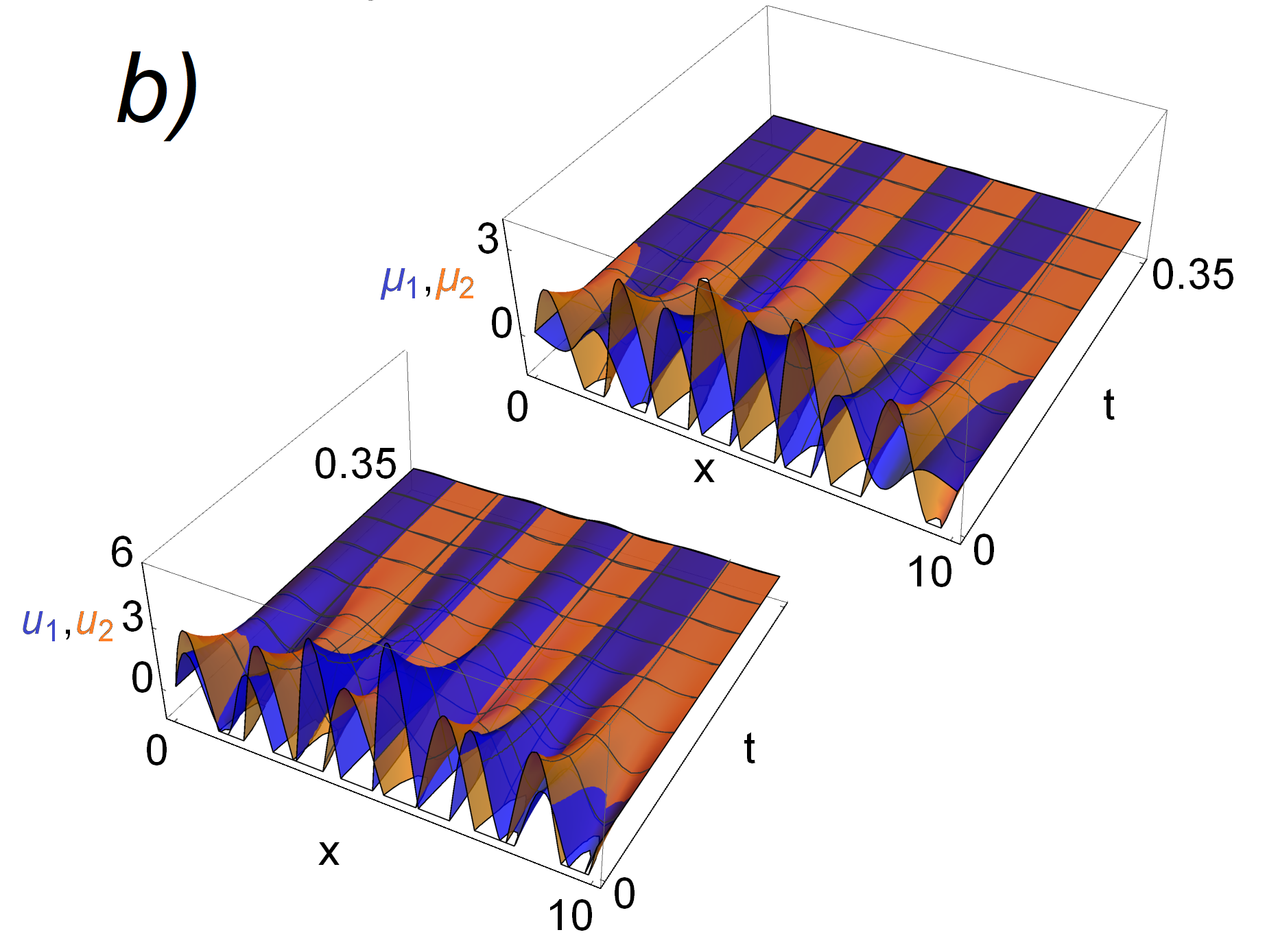}
\includegraphics[scale=0.20]{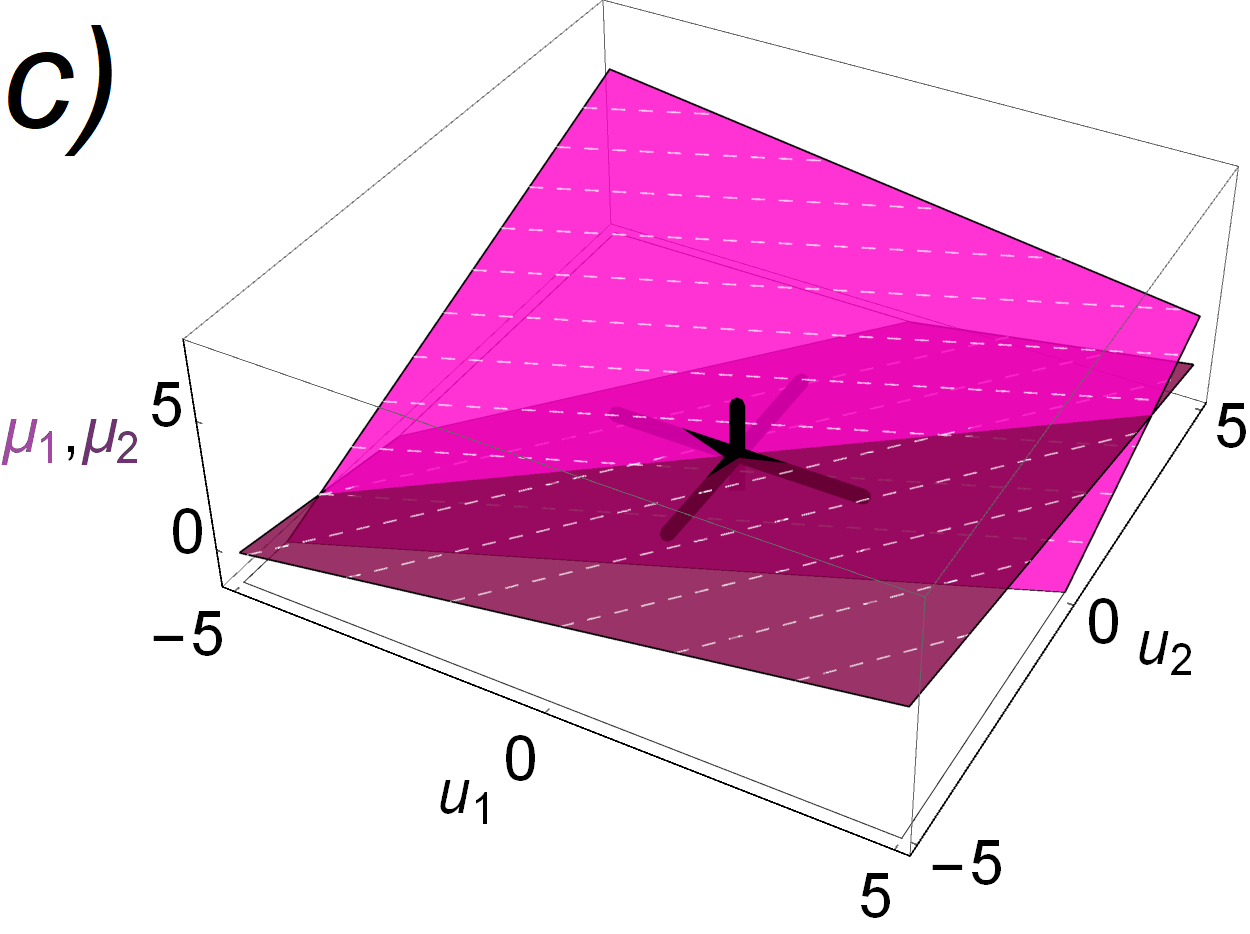}
\caption{Lyapunov function for a diffusive system. a) $\mathcal{V}$ as a function of time and the coordinates and evolution of some Fourier coefficients. b) Spatiotemporal concentrations and the chemical potentials. c) Coupling of chemical potentials.
}\label{fig:difuso}
\end{figure*}

In Fig. \ref{fig:difuso} we illustrate these concepts for the system in \eqref{eq:difusa} with the matrix of diffusion coefficients given in \eqref{eq:matD}. In Fig. \ref{fig:difuso}a, we plot the Lyapunov potential $\mathcal{V}(t)$ in terms of time and, in the inset, the decay of the three Fourier modes involved in our problem. As expected, at long enough times, all components reach the stationary concentration. These decays were obtained through numerical simulation of the diffusion equation. In  Fig. \ref{fig:reaction}b, we plot the solutions for the concentrations and the chemical potentials of both species, respectively. Finally, in Fig. \ref{fig:reaction}c we show the coupling between chemical potentials in terms of the concentrations.


\subsection{Lyapunov function and entropy production for non-equilibrium stationary states}

In the previous discussion, we have approached the case in which the homogeneous Dirichlet boundary conditions ensure the existence of a stationary state. Here, by relaxing this condition, we will show that the definition of the Lyapunov potential is robust enough to describe more interesting situations, such as those in which non-equilibrium conditions induce fluxes through the boundaries.

The starting point is the diffusion equation \eqref{eq:difusa}.  However, in the present case, for emphasizing the role of  external fluxes, we will consider \emph{non-homogeneous} Dirichlet boundary conditions given by
\begin{equation}\label{bb-non-hom}
\mathbf{n}(x=0,t)=\mathbf{n}^{left}, \quad \text{and} \quad \mathbf{n}(x=V,t)=\mathbf{n}^{right}.
\end{equation}
This means that if $\mathbf{n}^{left} \neq \mathbf{n}^{right}$, there is an effective flux crossing the domain.
In this case it is well known that the change of variables $\mathbf{u}(x,t)=\mathbf{n}(x,t) -\mathbf{n}^{ss}(x)$ will lead to the same homogeneous problem we have solved in this section for $\mathbf{u}(x,t)$. Therefore, we can conclude that, for stationary non-equilibrium boundary conditions like (\ref{bb-non-hom}), a general form of the Lyapunov potential for linear systems of the form \eqref{eq:difusa}  is 
\begin{equation}\label{V-nn}
\mathcal{V}(t) =\frac{1}{2}   \int_V |  \mathds{Q}^{-1}  \left (\mathbf{n} - \mathbf{n}^{ss} \right)   |^2 dv.
\end{equation}
\noindent where $ \mathbf{n}^{ss}$ is the stationary solution of  \eqref{eq:difusa} with the non-homogeneous boundary conditions (\ref{bb-non-hom}) given by
\begin{equation}
\mathbf{n}^{ss}(x)=\mathbf{n}^{left} +(x/V) \left[ \mathbf{n}^{right} - \mathbf{n}^{left}   \right].
\end{equation}
It should be noted that the form (\ref{V-nn}) is very general and depends on the boundary conditions imposed by each problem, so the properties of the dynamical system such as eigenvalues, initial and boundary conditions are reflected directly in our definition of thermodynamic properties.


\section{Reaction-diffusion systems}
The previous sections were devoted to obtaining Lyapunov functions for reactive and diffusive systems separately. Furthermore, its coherence with Gibbs free energy landscapes was demonstrated through the definition of a positive definite entropy production. We also show interesting relationships for the phenomenological Onsager coefficients and exhibit consistency in our definitions with relaxation to steady states under equilibrium and non-equilibrium conditions.

In the present section, we will show how to obtain a Lyapunov potential for linearized reaction-diffusion systems of the form
\begin{equation}\label{eq:rectiondiffusion}
\frac{d \mathbf{u}}{ dt }= \mathds{J} \mathbf{u}+\mathds{D}\nabla^2 \mathbf{u},
\end{equation}
\noindent with zero flux boundary conditions. The steps to obtain the Lyapunov potential are similar to the previous sections and, therefore, will be presented in summarized form.

One starts by proposing a solution of the form
\begin{equation}\label{eq:change3}
\mathbf{u}=\mathds{R}^{(0)} \mathbf{y}^{(0)}(t)+ \sum \limits_{m=1}^{\infty}\mathds{R}^{(m)} \, \mathbf{y}^{(m)}(t) \, \cos ( \kappa_m  x)
\end{equation}
where $\mathds{R}^{(m)}$ is the matrix whose columns are the eingevectors of $(\mathds{J}-\kappa_m^2 \mathds{D})$. The index $m$ reflects  Fourier mode dependency. Note that $\mathds{R}^{(0)}=\mathds{P}$ is the homogeneous part that corresponds to the chemical  reactions. This is solution of \eqref{eq:rectiondiffusion} if the functions $\mathbf{y}^{(m)}$ are uncoupled as 

\begin{equation}\label{eq:yrd}
\frac{d\mathbf{y}^{(m)}}{dt}= \Lambda^{(m)}  \mathbf{y}^{(m)}.
\end{equation}
where $\Lambda^{(m)}=\left[\mathds{R}^{(m)} \right]^{-1}(\mathds{J}-\kappa_m^2 \mathds{D})  \, \mathds{R}^{(m)}$ is a diagonal matrix whose entries, $\lambda_j(\kappa_m)\equiv \lambda_j^{(m)}$, with $j=1,2\ldots$ have non-positive real part if the system is linearly stable for all modes $m=0,1,2,\ldots$. Notice that the first term is proportional to  $\Lambda^{(0)}=\Lambda^{J}$ and is due to the reaction term. Using the definition of the reduced chemical potential given by Eq. \eqref{eq:mudifusa} (but starting at $m=0$ due to the boundary conditions), and of the Lyapunov potential \eqref{eq:lyapunov2}, then one finds

\begin{equation}\label{eq:mugeneral}
\hat{\boldsymbol{\mu}}^{(m)}= \left[ \mathds{R}^{(m)}\right]^{-T} \overline{\mathbf{y}}^{(m)} \cos (\kappa_m x).
\end{equation}
The orthogonality of the set of cosines is similar to  \eqref{eq:ortogonal}:
\begin{equation}\label{eq:ortocoseno}
\int_0^V \cos(\kappa_j x)  \cos(\kappa_k x)   dv=
\begin{cases}
 (V/2)  & \text{for } j,k=1,2,\ldots   \\
	 &  \text{and }j\neq k  \nonumber\\
    V & \text{for }  k=l=0.\\
   0 & \text{otherwise}.\\
 \end{cases}
\end{equation}
Using (\ref{eq:ortocoseno}), the following result is obtained for the Lyapunov function time derivative
\begin{equation}\label{eq:dvdtRD}
\frac{1}{V}\frac{d\mathcal{V}}{dt}= \left[ \overline{\mathbf{y}}^{(0)} \right]^{T} \Lambda^{(0)} \mathbf{y}^{(0)} + \frac{1}{2}\sum\limits_{m=1}^{\infty} \left[ \overline{\mathbf{y}}^{(m)} \right]^{T} \Lambda^{(m)} \mathbf{y}^{(m)}.
\end{equation}
The Eq. (\ref{eq:dvdtRD}) is negative as long as the eingenvalues $\lambda^{(m)}_k$ have negative real part for all modes $m$. Uncoupling the equations for $y_j^{(m)}$ in \eqref{eq:yrd}, integrating over time  and 
 including all eigenvalues $\lambda_k^{m}$ with null real part in the additive constant of the Lyapunov function, it can be shown that $\mathcal{V}$ is a combination of  \eqref{eq:lyapireacti} and \eqref{eq:vdifusa1}:

 \begin{equation}
 \frac{1}{V}\mathcal{V}(t)=\frac{1}{2} |\mathbf{y}^{(0)}|+\frac{1}{4}  \sum\limits_{m=1}^{\infty}  \left| \mathbf{y}^{(m)}\right|^2.
 \end{equation}
To write this equation in terms of the original concentrations, we use the temporal dependent Fourier coefficients  $\mathbf{a}_k(t)$ defined as usual through
 
 \begin{equation}
 \mathbf{a}_l (t)=\frac{2}{V}\int _V \mathbf{u}(x,t)  \cos (\kappa_l x) dv,
 \end{equation}
 in order to obtain
 
 \begin{equation}\label{eq:lyapiRD}
 \mathcal{V}(t)=\frac{V}{2} \left| \left[\mathds{R}^{(0)}\right]^{-1} \frac{\mathbf{a}_0 (t)}{2} \right|^2 +\frac{V}{4} \sum\limits_{m=1}^{\infty} \left| \left[ \mathds{R}^{(m)}\right]^{-1} \mathbf{a}_m (t) \right|^2.
 \end{equation}
 This expression shows that the function $\mathcal{V}$ is positive definite and attaches its minimum value when all the Fourier coefficients are null, \emph{i.e.} at the stationary solution of \eqref{eq:rectiondiffusion}, $\mathbf{u}^{ss}=0$  As before, in principle fixed points of the reaction  $\mathbf{n}_0$ other than $\mathbf{0}$, as well as other boundary conditions  leading to different stationary solutions can be considered in this methodology by making the change of variables $\mathbf{u}=\mathbf{n}(x,t)-\mathbf{n}_0+\mathbf{n}^{ss}(x)$.

\subsection{Thermodynamic consequences}
The first order approximation of the chemical potential near the stationary solution can be found from \eqref{eq:mudifusa}, \eqref{eq:yrd} and the orthogonalization condition  \eqref{eq:ortocoseno} as

\begin{equation}\label{eq:potentialR}
\frac{\Delta\boldsymbol{\mu}}{ \mathcal{R} T } \cong \hat{\boldsymbol{\mu}}=  \mathds{L}^{(0)} \frac{\mathbf{a}_0(t)}{2}+\sum\limits_{m=1}^{\infty}  \mathds{L}^{(m)} \mathbf{a}_m(t) \cos (\kappa_m x),
\end{equation}
with 
\begin{equation}
\mathds{L}^{(m)}=\left[\mathds{R}^{(m)}\right]^{-T} \left[\overline{\mathds{R}}^{(m)}\right]^{-1}.
\end{equation}
For the chemical-reaction term, we can identify the same matrix of proportionality $ \mathds{L}^{(0)}=\mathds{L}^{J}$. For the diffusive part, the positive definite matrices  $ \mathds{L}^{(m)}$ with $m=1,2,\ldots$ generally depend on the wavenumber $\kappa_m$.

In the same way as for pure chemical reactions, the chemical potential of one species is not independent of the other species in irreversible reactions. Here we show that when diffusion in the reactor is included, the \emph{effective} chemical potential of each species cannot be taken only as the one that results from the sum of the diffusion and chemical processes, since in general:
\begin{equation}
\mathds{L}^{(m)} \neq ( \mathds{L}^J +\mathds{L}^D).
\end{equation}
The above relationship has a definite physical meaning in terms of the coupling of the different processes in the system. It tells us that when in an initially homogeneous mixture of species, the mixing process stops and the species are free to evolve and diffuse, then this evolution occurs in such a way that a rearrangement of flows and forces occurs. This rearrangement is not necessarily the addition of a new term in the chemical potential.

\begin{figure*}[btp]
\centering
\includegraphics[scale=0.20]{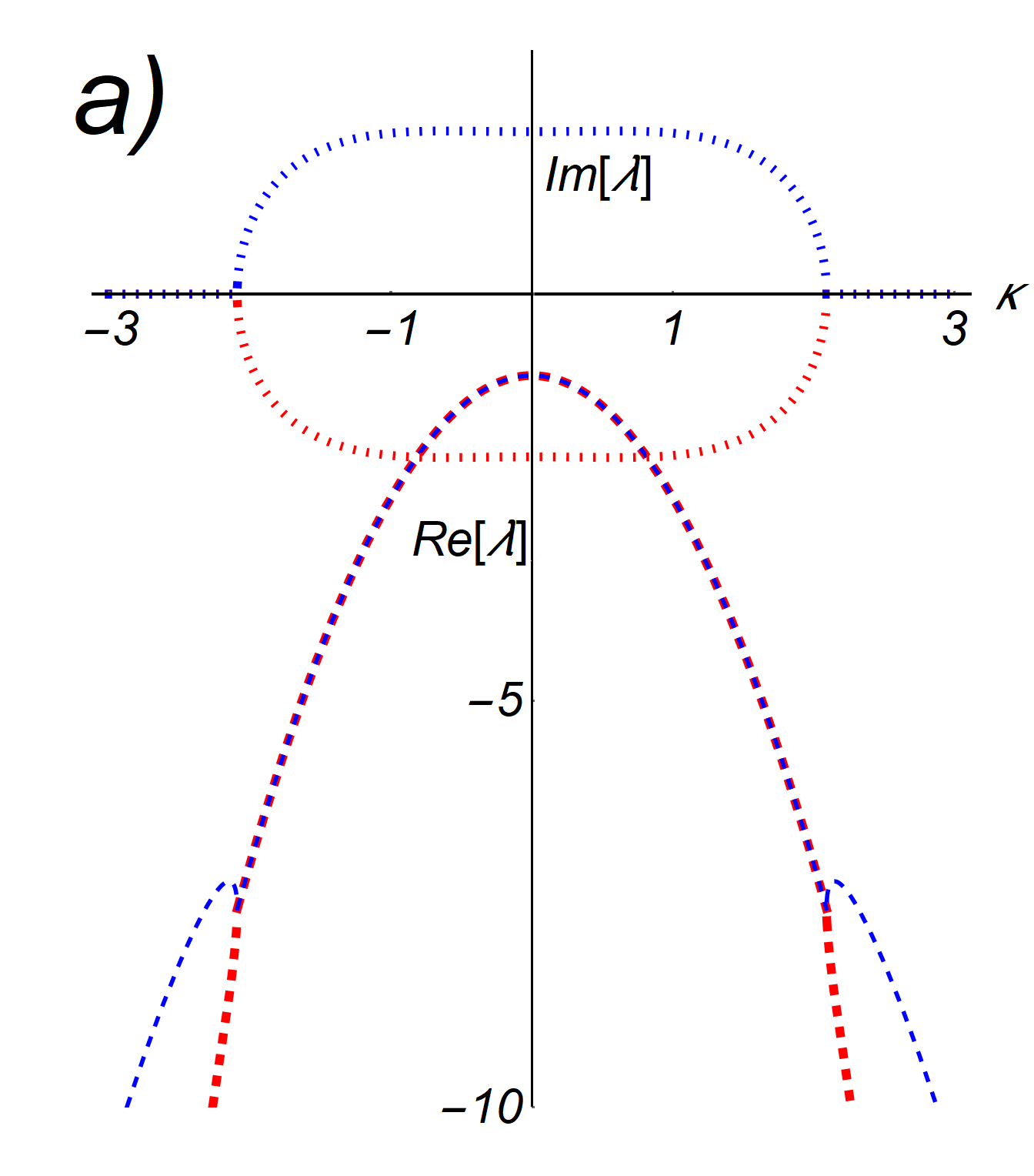}
\includegraphics[scale=0.15]{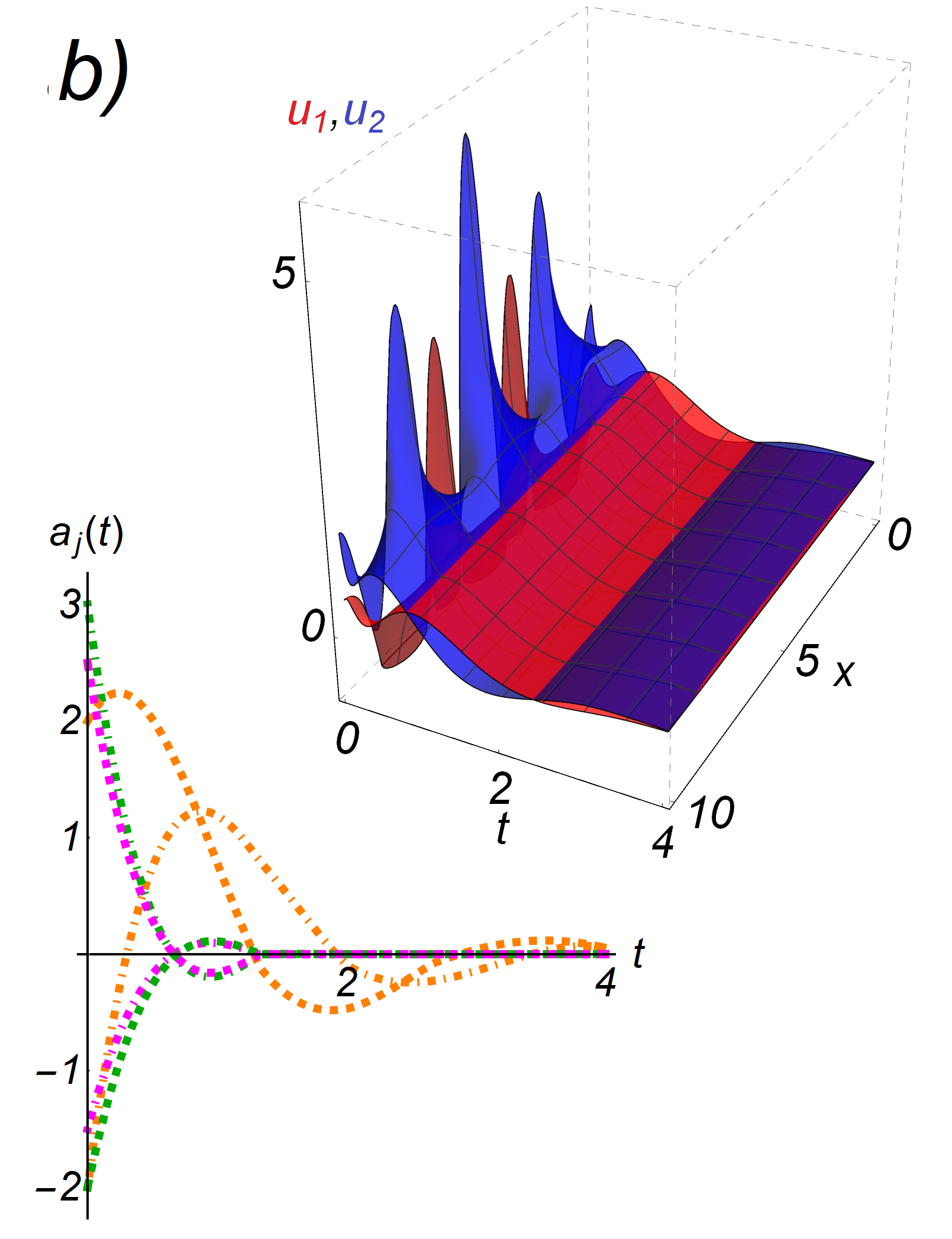}
\includegraphics[scale=0.30]{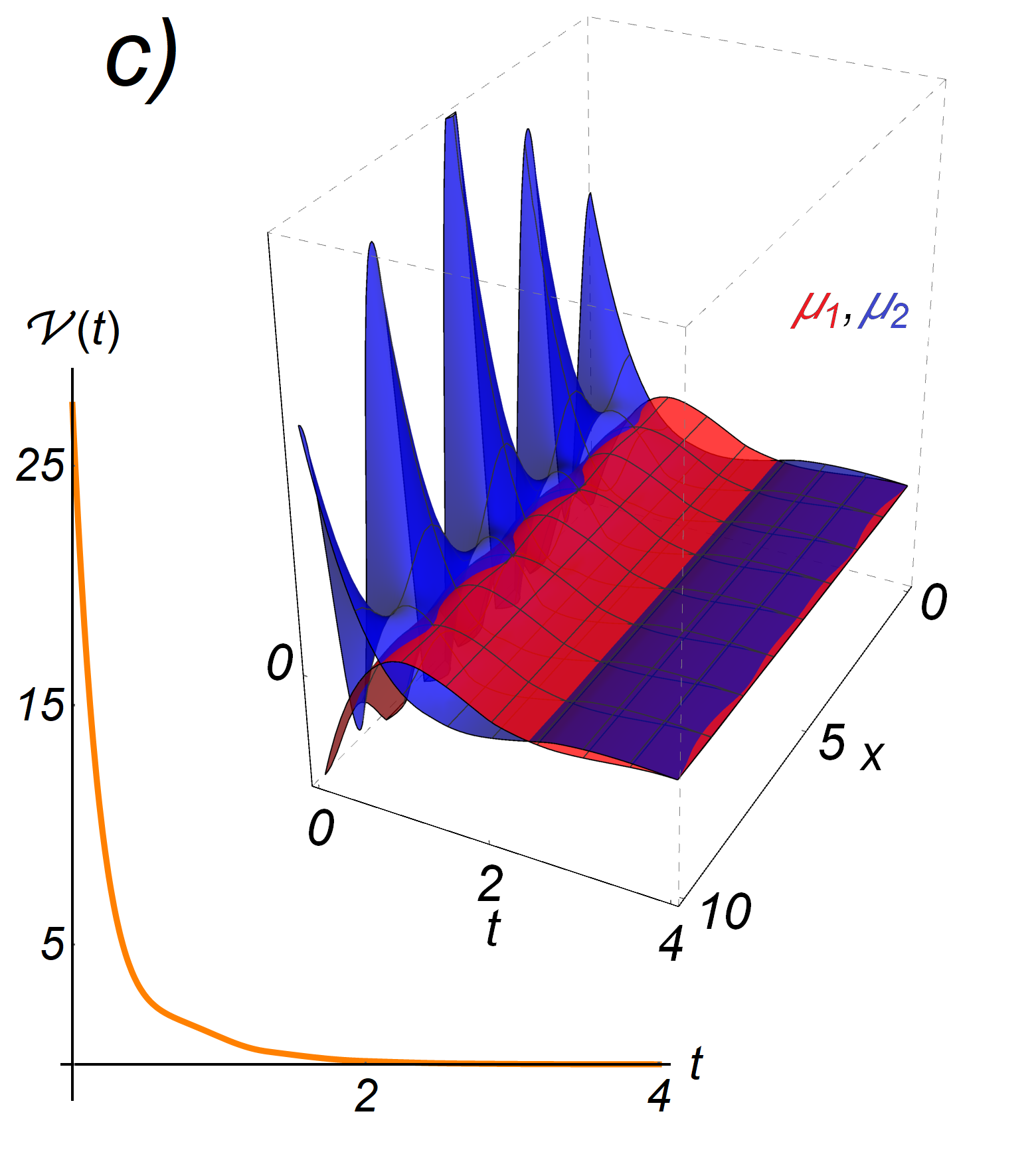}
\caption{Linearly stable reaction-diffusion system. a) Dispersion relation b) Spatio-temporal profile of the solution and its respective evolution of the Fourier modes. c) Temporal evolution of the Lyapunov functional $\mathcal{V}$ and the chemical potential from where it comes.
}\label{fig:rdimage}
\end{figure*}

The above observation is illustrated in Fig. \ref{fig:rdimage} for a reaction diffusion system of the form \eqref{eq:rectiondiffusion} with the two matrices previously given in \eqref{eq:jota} and \eqref{eq:matD}. 
 To obtain the temporal evolution of the potential $\mathcal{V}$, we solve \eqref{eq:rectiondiffusion} numerically and compute the Fourier modes $\mathbf{a}_k(t)$. As illustrated in Fig. \ref{fig:rdimage}b, the solution has some transient temporal oscillations combined with the homogenizing effect of diffusion. Inserting these Fourier modes in \eqref{eq:lyapiRD}, we obtain the energy decay in the system illustrated in Fig. \ref{fig:rdimage}c, as well as the interaction between both chemical potentials. This example illustrates the importance of initial and boundary conditions in reconstructing the thermodynamic properties of the reaction-diffusion system. Fig. \ref{fig:rdimage}a shows the dispersion relation and corroborates that our formalism is valid as long as the eigenvalues remain in the stable region.

\section{Lyapunov function and entropy production rate. Physical considerations }

This section will briefly discuss the consistency of our results with some fundamental physical aspects such as entropy in an open and irreversible system, the Glansdorff-Prigogine criterion, and the spectral decomposition of entropic production. For the sake of simplicity, we will focus only on the chemical reactions part, although the diffusion process can easily be included.

\paragraph*{Difference between stationary and equilibrium solutions}

We begin our discussion by specifying the difference between stationary and equilibrium states by considering a reactor in which the following general reaction occurs:
\begin{center}
\ce{$\sum$R$_i$ + X + Y -> $\sum$P$_j$} + X + Y.\\
\end{center}
Here $R_i$ states for reagents and $P_j$ for products. We will assume that the passage from reactants to products occurs through a series of chemical transformations represented by the following mechanism of one-step irreversible reactions
\begin{equation}\label{eq:reacciones}
\begin{array}{c}
\qquad\qquad\ce{R1 ->[\ce{k1}] X + P1}\\
\ce{R2 + X ->[\ce{k2}] P2}\\
\qquad \ce{R3 + Y ->[\ce{k3}] X + P3}\\
\qquad\qquad \ce{R4 ->[\ce{k4}] Y + P4}\\
\ce{R5 + Y ->[\ce{k5}]  P5}\\
\qquad \ce{R6 + X ->[\ce{k6}] Y + P6}.\\
\end{array}
\end{equation}

\noindent In the general case, all species concentrations, including catalysts (X and Y) and reactants and products, will depend on time and therefore will produce time-dependent internal production of entropy. However, in a chemostat reactor, it is commonly assumed that the latter species, $ R_i $ and $ P_j $, can be kept at a constant concentration. Therefore, internal changes are fully described by the law of mass action applied to $ x $ and $ y $ (lower case is for concentrations). This leads to
\begin{equation}\label{eq:system2}
\begin{array}{cl}
\dot{x}&=a_1  - (a_2 + a_6) x+ a_3 y,\\
\dot{y}&= a_4 + a_6 x-(a_3+a_5) y,\\
\end{array}
\end{equation}
\noindent that clearly shows the linearity of the equations since the combinations $ a_i = k_i r_i $ are constant. Therefore, the rate of entropy production in a chemostat can be quantified using only catalytic species such as

\begin{equation}\label{eq:1}
\sigma=\frac{d_is}{dt}=  -\frac{1}{T} \sum \limits_{k=1}^{2}  \mu_k \frac{d n_k}{ dt }= -\frac{1}{T} \left[  \mu_x \frac{d x}{ dt }+ \mu_y \frac{d y}{ dt }  \right].
\end{equation}
\noindent Since  the total change of entropy density inside the reactor is $ds/dt =-\nabla \cdot \mathbf{J}_s +\sigma$, then a stationary state occurs when $ds/dt=0$. Here $\mathbf{J}_s$ is the flux of entropy through the boundaries of the reactor. This implies that in the stationary state:

\begin{equation}
\oint_A  \mathbf{J}_s\cdot \mathbf{dA}=\int_V\sigma\, dv,
\end{equation}
\noindent  which means that the entropy produced within the reactor is offset by that exchanged with the surroundings.

Therefore, different types of stationary states can occur: 1)  
when $ \mathbf{J}_s=0$ there is state of equilibrium  since there is no exchange of entropy with the outside and therefore $\sigma=0$; 2) when $\nabla \cdot \mathbf{J}_s=0$ but $\mathbf{J}_s \neq 0$, for example, the case $\mathbf{J}_s=$cte. In this case, there is an external flow of material but still $\sigma=0$ and it constitutes a kind of stationary non-equilibrium state. 3) 
When neither $\sigma$ nor $\mathbf{J}_s$ are zero but still $\nabla \cdot \mathbf{J}_s =\sigma$,  a steady non equilibrium state is attached. In this work, we analyze the first two cases, and the third is still pending \cite{ledesmafuturo}.

\paragraph*{Affinity and velocities of reaction}
We want to emphasize that our formalism allows us to include open and irreversible systems for which the detailed balance is not necessary to achieve a stationary state. In a complementary way, the presented Lyapunov formalism establishes that the differences in chemical potentials and concentrations are the new flows and forces. However, the same results are obtained if we write the entropy production rate in terms of the velocities of reaction  $ v_ {\rho} $, and the affinities $ A_\rho $. Here, for the mechanism \eqref{eq:reacciones} we have

\begin{equation}\label{eq:entropy2}
\frac{d_i s}{dt}=  \frac{1}{T} \sum \limits_{\rho=1}^{6} A_{\rho} v_{\rho},
\end{equation}

 \noindent Here, the velocities are $v_\rho=d\xi_\rho/dt$ with $\xi_\rho$ the reaction coordinate of the $\rho$-th reaction. They are given in accordance with mass law action by 
\begin{equation}
\begin{array}{ccc}
\frac{d\xi_1}{dt}=a_1, & \frac{d\xi_2}{dt}=a_2 x,& \frac{d\xi_3}{dt}=a_3 y, \\
\frac{d\xi_4}{dt}=a_4,&\frac{d\xi_5}{dt}=a_5 y,&\frac{d\xi_6}{dt}=a_6 x. \\
\end{array}
\end{equation}
Using Eq. \eqref{eq:entropy-explicit2}, the affinities are 
\begin{equation}
A_{\rho}=-\sum_{k=1}^2 \nu_{\rho\,k}\mu_k=-\mathcal{R}T \sum_{k=1}^2 \nu_{\rho\,k}  \sum\limits_{l=1}^2 \mathds{L}_{kl}  (n_l-n_{l0}).
\end{equation}
\noindent where the stoichiometric matrix with coefficients $\{\boldsymbol{\nu}\}_{\rho\,k}$ for the reactions we are considering in \eqref{eq:system2} is 
\begin{equation}
\boldsymbol{\nu}=\left(\begin{array}{cccccc}
+1 & -1& +1& 0 & 0& -1\\
0& 0& -1& 1 & -1& +1
\end{array}\right)^T.
\end{equation}
Note that the reaction velocities include only forward reaction rates since the steps are irreversible. This means that the hypothesis used in other works \cite{yoshimura2020information,
falasco2018information}, that all systems are to some extent reversible, is not necessary for our formalism.

\paragraph*{Glansdorff-Prigogine criterion} 
The Glasndorf-Prigogine criterion establishes the evolution of the state of a thermodynamic system. Here, we will show that it holds under our formalism \cite{glansdorff1971thermodynamic}. To show this we consider the entropy production rate as $P=\int {d_is}/{dt} dV$. Then, identifying the forces with $X_k=-\Delta \mu_k/T$ and the fluxes as $F_k=dn_k/dt$, we can define in the usual way \cite{nicolis1977self}:
\begin{equation}
\frac{dP_X}{dt}=-\frac{1}{T}  \sum\limits_{k=1}^{2} \int_V \frac{d\mu_{k}}{dt} \frac{dn_k}{dt}dV.
\end{equation}
\noindent Using the chain rule we obtain

\begin{equation}
\frac{dP_X}{dt}=-\frac{1}{T} \sum\limits_{j,k=1}^2  \int _V  \frac{dn_j}{dt}\left( \frac{\partial \mu_ {k}}{\partial n_j}\right)\frac{dn_k}{dt} dV.
\end{equation}
\noindent Eq. \eqref{eq:entropy-explicit2} leads to the identification
$\{\mathds{L}^{J}\}_{k\,j}=(1/\mathcal{R}T) \partial \mu_k/ \partial n_j$, and therefore:

\begin{equation}\label{eq:glando}
\frac{dP_X}{dt}=- \mathcal{R} \int_V \frac{d \mathbf{n}}{dt}^T \mathds{L}^{J} \frac{d \mathbf{n}}{dt}\,dV \leq 0.
\end{equation}
Here, the inequality follows from the fact that the matrix $\mathds{L}^{J}$ is positive definite. Besides, using  \eqref{eq:lj}, we can write \eqref{eq:glando} as

\begin{equation}\label{eq:px}
\frac{dP_X}{dt}=-\mathcal{R}\int_V \left| \mathds{P}^{-1} \frac{d\mathbf{n}}{dt} \right|^2 dV.
\end{equation}
From this last relation, it follows that the entropy production rate reaches its minimum value in the stationary state, which is the fixed value of the concentrations. Something similar occurs with the diffusion and RD processes. Note the similarity of \eqref{eq:px} with the relation in \eqref{eq:lyapireacti} for the Lyapunov function.

\paragraph*{Entropy decomposition in Lyapunov spectra } 

Finally, from \eqref{eq:change}, \eqref{eq:entropy} and \eqref{eq:sumadvt}
we obtain

\begin{equation}\label{eq:nieto}
\sigma=- \mathcal{R} \sum \limits_k^{N} \lambda_k \left|\sum \limits_l^{N} \mathds{P}^{-1}_{kl} \mathbf{u}_l \right|^2.
\end{equation}
\noindent This last result is in agreement with the hypothesis that the entropy production is approximately proportional to the spectra of Lyapunov exponents, which was studied in Refs. \cite{hoover1994second,betancourt2016entropy,gaspard2007time}.  Our formalism tests this fact strictly for the linear case. The study of the validity of expression \eqref{eq:nieto} in the nonlinear case remains a matter for future studies. Note that the entropy grows proportionally to the Lyapunov exponents (in absolute value) and stops growing in the steady state where $\mathbf{u}=\mathbf{0}$.

\section{Discussion and conclusions}
In this work, we have presented a mathematical treatment of the irreversible linear reaction and reaction-diffusion systems in which the Lyapunov function is proportional to the internal free energy and its time derivative proportional and of opposite sign to the entropy production rate. The fundamental strategy was to use the fact that the thermodynamic properties of a real system are reflected in the solution of the dynamic equations that represent its temporal evolution. From this starting point, we have proposed a method to infer the expressions of the chemical potentials in such a way as to guarantee the second law of thermodynamics  (positive entropy production and energy minimization).

 Our methodology inserts well-defined properties of dynamical systems, such as eigenvalues, eigenvectors, initial and boundary conditions, and stationary solutions, directly into the definitions of chemical potentials, energy, and entropy. This method allows us to re-understand the interaction between flows and forces that, proceeding in the usual way for reversible reactions, would lead to very restricted conditions that do not allow some observable behaviors to be included \cite{chen2019cross}. For example, although traditional approaches certainly allow us to include irreversibility as a limit case \cite{yoshimura2020information,falasco2018information}, the detailed balance assumption leaves out most of the usual reaction-diffusion models that proceed, in general, following irreversible mechanisms \cite{ge2012landscapes}.

 We have restricted our analysis to linearly stable systems, that is, those whose dispersion relationship arising from a linear approach gives eigenvalues with a negative real part. This property is entirely consistent with the hypothesis of linearity between flows and forces and helps to visualize the central role of non-diagonal terms in the couplings between the chemical potentials of the different species in the reaction mechanism. For example, we have illustrated this coupling for the interaction of reaction and diffusion processes. In general, it seems that the chemical potentials involved in a reaction-diffusion system cannot be obtained simply by knowing the information of the two processes separately.

Regarding the apparent restriction of our approach to the linearity of the dynamical system, we believe that our formalism can be generalized for reaction schemes more general than those given in \eqref{eq:system2}, as long as the results are applied locally for each stable fixed point \cite{shapiro1979possibility}. The  Hartman-Grobman theorem guarantees that nonlinear systems remain stable in a neighborhood of an asymptotic stable fixed point. In this case, the Lyapunov function found by us is only an approximation of the energy near the stable fixed value of the concentrations when nonlinearities are small or negligible \cite{boros2019complex,auchmuty1975bifurcation}. 
  
Regarding the linear proportionality between flows and forces, the question is more delicate. For the diffusion process, the linear relationship between diffusion flux and concentration gradient is given by Fick's diffusion laws and appears to be valid for a wide range of concentrations. On the contrary, in the field of chemical reactions, the linear proportionality between chemical flows and forces seems strictly valid only near the equilibrium \cite{nicolis1998probabilistic,zarate2017effect,caceres2017close}.  This question constitutes an advantage for our formalism, where this relation of proportionality emerges naturally for linear systems. However, it will be interesting to study how nonlinear terms in chemical kinetics modifies this relationship. 

We believe that our approach can be generalized to nonlinear systems. This will require a method that allows us to consider the interaction of nonlinear terms in a manageable way, such as the multiscales method near a Turing or Hopf bifurcation \cite{kumar2020energetic,ledesma2019spatio}. This will allow us to understand the thermodynamic properties of dissipative structures in terms of their observable properties such as amplitude and wavenumber, in contrast to traditional approaches where this relationship is not always clear \cite{ledesmafuturo}. 
The strategy coincides with the current one, \emph{i.e.}, starting from an established chemical reaction mechanism capable of presenting dissipative structures, calculating a Lyapunov function and, from this, also the internal energy, the entropy change and the chemical potentials.

\section*{Acknowledgments}
 We thank the Divisi{\'o}n de Ciencias B{\'a}sicas e Ingenier{\'i}a of UAM-Iztapalpa for financial support under the \emph{Programa Especial de Apoyo a Proyectos de Docencia e Investigaci{\'o}n, 2021}.

		
%


\setcounter{equation}{0}
\renewcommand\theequation{A.\arabic{equation}}

\section*{Appendix: Numerical examples}
To illustrate these ideas with numerical examples, we consider three examples of these processes: a reactive system, a cross-diffusion process, and finally, an RD system, all in a one-dimensional volume. Consult the main text for its physical and mathematical interpretation.

\subsection*{Example of purely reactive System}
We first  consider the chemical system of the form  \eqref{eq:linear} with Jacobian in $\mathbf{n_0}$ given by 

\begin{equation}\label{eq:jota}
\mathds{J}=\frac{1}{9}\left(
\begin{array}{cc}
 -15 & 18 \\
 -20 & -3 \\
\end{array}
\right),
\end{equation}
whose eigenvalues are $-1+2 i,-1-2 i$ and their matrix of eigenvectors is 

\begin{equation}
\mathds{P}=\frac{1}{10}
\left(
\begin{array}{cc}
 3-9i & 2+9i\\
 10 & 10 \\
\end{array}
\right).
\end{equation}
The real solutions of this system are 

\begin{equation}
\begin{array}{cl}
 u_1(t)=&\frac{1}{3} e^{-t} [3 c_1\cos (2 t)-( c_1-3 c_2) \sin (2 t)],\\
  u_2(t)=&\frac{1}{9} e^{-t} [(3 c_2-10 c_1) \sin (2 t)+9 c_2\cos (2 t)].\\
\end{array}
\end{equation} 
\noindent In this case, the Lyapunov function from \eqref{eq:lyapireacti} is 
\begin{equation}
\mathcal{V}(t)=\frac{25} {81}\left[  \left( u_1-\frac{3}{10}u_2   \right)^2 + \frac{81}{100} u_2^2  \right],
\end{equation}
and the chemical potential is given by \eqref{eq:entropy-explicit2} as 

\begin{equation}
\Delta\boldsymbol{\mu}=-
\frac{1}{405}\left(
\begin{array}{cc}
10 & -3 \\
 -3 & 15 \\
\end{array}
\right)
\left(
\begin{array}{c}
u_1 \\
u_2 \\
\end{array}
\right).
\end{equation}
where the components of $\mathbf{u} \approx \mathbf{n}-\mathbf{n}_0$ represent the differences of concentrations respect to the fixed point.

\subsection*{Example of purely diffusive System}
For the pure diffusive system in the form \eqref{eq:difusa}, we consider a diffusion matrix 

\begin{equation}\label{eq:matD}
\mathds{D}=\frac{1}{16}
\left(
\begin{array}{cc}
18 & -4 \\
 -7& 30 \\
\end{array}
\right),
\end{equation}
with homogeneous Dirichlet boundary conditions in a one dimensional domain of length $V=10$ and  initial conditions with only three Fourier terms (for simplicity) given by 

\begin{equation}
\begin{array}{cl}
u_1(x,0)&= \sin \left(\kappa_4 x \right)-2 \sin \left( \kappa_5 x \right)+3 \sin \left(\kappa_6 x \right),\\
u_2(x,0)&=-\sin \left(\kappa_4 \right)+3 \sin \left(\kappa_5 x \right)+ \sin \left( \kappa_6 x \right).
\end{array}
\end{equation}
For this problem, the matrix of conversion is

\begin{equation}
\mathds{Q}=\frac{1}{7}
\left(
\begin{array}{cc}
 -2 & 14 \\
 7 & 7 \\
\end{array}
\right),
\end{equation}
from which is deduced that the Lyapunov function for the problem \eqref{eq:lyapidif} is 

\begin{equation}
\mathcal{V}(t)=  \frac{49 }{256}  \int\limits_0^{10}\left[  \left ( u_1- \frac{6}{7}u_2 \right)^2 +\frac{64}{49}u_2^2 \right] dv.
\end{equation}

\subsection*{Example of linear RD system}
Now let us consider the reaction diffusion system of the form \eqref{eq:rectiondiffusion} with the matrices $\mathds{J}$ and $\mathds{D}$ given by \eqref{eq:jota}
and \eqref{eq:matD}.  We use zero flux boundary conditions and initial conditions with just three modes (for illustrative purposes) as:

\begin{equation}
\begin{array}{cl}
u_1(x,0)=&-1+3 \cos \left(\kappa_3 x\right)-\frac{3}{2} \cos \left( \kappa_4 x \right),\\
u_2(x,0)=&1-2 \cos \left(\kappa_3 x\right)+\frac{5}{2} \cos \left(\kappa_4 x \right).
\end{array}
\end{equation}
\noindent For this problem, the dispersion relationship is $\lambda ^2+\left(3 \kappa ^2+2\right) \lambda +(2 \kappa ^4+229 \kappa ^2/72+5)=0$. It provides complex eigenvalues with real part negative and therefore the solutions are linearly stable, see Figure \ref{fig:rdimage}a. The matrix of conversion is 

\begin{equation}
\mathds{R}^{(m)}=\frac{1}{320-63 \kappa ^2}
\left[
\begin{array}{cc}
 -6 \left(9 \kappa ^2+2 \Delta -16\right) & 6 \left(-9 \kappa ^2+2 \Delta +16\right) \\
 320-63 \kappa ^2 & 320-63 \kappa ^2 \\
\end{array}
\right]
\end{equation}
with $\Delta=36 \kappa ^4-26 \kappa ^2-576$, the discriminant of the dispersion relation.

%



\end{document}